\documentclass[twocolumn]{aa}

\usepackage{graphicx}
\usepackage{txfonts}
\usepackage{xspace}

\begin{document}

\newcommand{\hess}{H$\cdot$E$\cdot$S$\cdot$S\xspace}
\newcommand{\mmm}{M\,87\xspace}


  \title{Is the Giant Radio Galaxy \mmm a TeV Gamma-Ray Emitter?}

  \author{
F.~Aharonian\inst{1},
A.~Akhperjanian\inst{7},
M.~Beilicke\inst{4},
K.~Bernl\"ohr\inst{1},
H.-G.~B\"orst\inst{5},
H.~Bojahr\inst{6},
O.~Bolz\inst{1},
T.~Coarasa\inst{2},
J.L.~Contreras\inst{3},
J.~Cortina\inst{10},
S.~Denninghoff\inst{2},
M.V.~Fonseca\inst{3},
M.~Girma\inst{1},
N.~G\"otting\inst{4},
G.~Heinzelmann\inst{4},
G.~Hermann\inst{1},
A.~Heusler\inst{1},
W.~Hofmann\inst{1},
D.~Horns\inst{1},
I.~Jung\inst{1},
R.~Kankanyan\inst{1},
M.~Kestel\inst{2},
A.~Kohnle\inst{1},
A.~Konopelko\inst{1},
H.~Kornmeyer\inst{2},
D.~Kranich\inst{2},
H.~Lampeitl\inst{4},
M.~Lopez\inst{3},
E.~Lorenz\inst{2},
F.~Lucarelli\inst{3},
O.~Mang\inst{5},
H.~Meyer\inst{6},
R.~Mirzoyan\inst{2},
A.~Moralejo\inst{3},
E.~Ona-Wilhelmi\inst{3},
M.~Panter\inst{1},
A.~Plyasheshnikov\inst{1,8},
G.~P\"uhlhofer\inst{1},
R.~de\,los\,Reyes\inst{3},
W.~Rhode\inst{6},
J.~Ripken\inst{4},
G.~Rowell\inst{1},
V.~Sahakian\inst{7},
M.~Samorski\inst{5},
M.~Schilling\inst{5},
M.~Siems\inst{5},
D.~Sobzynska\inst{2,9},
W.~Stamm\inst{5},
M.~Tluczykont\inst{4},
V.~Vitale\inst{2},
H.J.~V\"olk\inst{1},
C.~A.~Wiedner\inst{1},
W.~Wittek\inst{2}}

\institute{
        Max-Planck-Institut f\"ur Kernphysik,
        Postfach 103980, D-69029 Heidelberg, Germany
        \and
        Max-Planck-Institut f\"ur Physik, F\"ohringer Ring 6,
        D-80805 M\"unchen, Germany
        \and
        Universidad Complutense, Facultad de Ciencias
        F\'{\i}sicas, Ciudad Universitaria, E-28040 Madrid, Spain
        \and
        Universit\"at Hamburg, Institut f\"ur
        Experimentalphysik, Luruper Chaussee 149,
        D-22761 Hamburg, Germany
        \and
        Universit\"at Kiel, Institut f\"ur Experimentelle und
        Angewandte Physik,
        Leibnizstra{\ss}e 15-19, D-24118 Kiel, Germany
        \and
        Universit\"at Wuppertal, Fachbereich Physik,
        Gau{\ss}str.20, D-42097 Wuppertal, Germany
        \and
        Yerevan Physics Institute, Alikhanian Br. 2, 375036
        Yerevan, Armenia
        \and
        On leave from  
        Altai State University, Dimitrov Street 66, 656099 Barnaul, Russia
        \and
        Home institute: University Lodz, Poland
        \and
        Now at Institut de F\'{\i}sica d'Altes Energies,
        UAB, Edifici Cn, E-08193, Bellaterra (Barcelona), Spain
        }

  \offprints{
    N.~G\"otting, M.~Tluczykont
    \\
    {\scriptsize \email{ Niels.Goetting@desy.de, Martin.Tluczykont@desy.de}}
    }

  \date{Received / Accepted}

  \abstract{
    For the first time an excess of photons above 
    an energy threshold of 730\,GeV from the giant radio galaxy
    \mmm  has been measured at a significance level above 4\,$\sigma$.
    The data have been taken during the years 1998 and 1999 with the
    HEGRA stereoscopic system of 5 imaging atmospheric Cherenkov
    telescopes. The excess of $107.4 \pm 26.8$ events above 730\,GeV
    corresponds to an integral flux of 
    3.3\,\% of the Crab flux
    or $N_\gamma(E > 730\,\mbox{GeV}) = (0.96 \pm 0.23) \times
    10^{-12}$\,phot.\,cm$^{-2}$s$^{-1}$. \mmm is located at the 
    center of the Virgo cluster of galaxies at a relatively small redshift
    of $z = $ 0.00436 and is a promising candidate among the class of giant
    radio galaxies for the emission of TeV $\gamma$-radiation. The detection
    of TeV $\gamma$-rays from \mmm\ --~if confirmed -- would establish
    a new class of extragalactic source in this energy 
    regime since all other AGN detected to date at TeV energies are
    BL Lac type objects.
 
    }

  \authorrunning{Aharonian et al.}
  \maketitle



\section{\label{chapter1}Introduction}

Active Galactic Nuclei (AGN) are believed to contain as a central ``engine'' 
a supermassive black hole which causes the development of large scale jets.
Extragalactic TeV $\gamma$-ray emission 
has been observed so far from AGN only of the BL Lac type, i.\,e.~objects 
ejecting matter in a jet oriented very close to the observer's line of sight. In BL 
Lacs, TeV photons are commonly believed to originate in the relativistic 
jets, most popularly due to inverse Compton scattering.
The well studied objects Mkn\,421 (redshift 
$z =$~0.030) 
and Mkn\,501 ($z =$~0.034) 
belong to this type of TeV $\gamma$-ray emitters.
Recently, the BL Lac type objects 1ES\,1959+650 ($z =$~0.047) (\cite{7ta};
\cite{hegra_1ES1959_2002}) and the much more distant H\,1426+428 ($z $ =
0.129) (\cite{horan_2002}; \cite{hegra_1426}) have also been established as TeV 
$\gamma$-ray emitters.
However, other types of AGN, e.\,g.~giant radio galaxies,
also show relativistic mass outflows, though, in contrast 
to BL Lac type objects, under large viewing angles. Amongst these the 
nearby radio galaxy \mmm has been speculated to be a powerful accelerator of cosmic
rays (including the highest energy particles observed in the universe,
see e.\,g.~\cite{ginzburg_m87}; \cite{biermann_uhecr_m87}).
\mmm has been targeted with the HEGRA Cherenkov telescopes as
one of the prime candidates for TeV $\gamma$-ray emission from this class
of objects. 

The elliptical galaxy \mmm 
(right ascension $\alpha_{\mbox{\tiny J2000.0}}$ : 
   $12^{\mbox{\tiny hr}}30^{\mbox{\tiny m}}$49.4$^{\mbox{\tiny s}}$,
declination~$\delta_{\mbox{\tiny J2000.0}}$ : 
   $+12^\circ23^\prime28^{\prime\prime}$,
redshift $z = 0.00436$)
has an optical extension
of $8.3^\prime \times 6.6^\prime$ (\cite{m87_position}) with a large radio
halo of $16^\prime \times 12^\prime$ (\cite{m87_cameron_1971}).
\mmm contains a supermassive black hole with a mass
$M_{\mbox{\tiny BH}} \approx 2 \mbox{--} 3 \times 10^9\,\mbox{M}_\odot$
(\cite{M87_harms}).
The power of the non-thermal jet is estimated to be as high as
a few $10^{44}\,\mbox{erg\,s}^{-1}$ (\cite{m87_owen_2000}).
The angle of the \mmm jet axis
to the line of sight was determined to be $30^\circ$ -- $35^\circ$
(\cite{M87_bicknell}). \mmm is located in the central region of the Virgo
cluster of galaxies, which itself is another interesting site for particle
acceleration (e.\,g.~\cite{virgo_cr_voelk}).

The VERITAS collaboration has targeted \mmm
with the Whipple 10\,m
Cherenkov telescope in the years 2000 and~2001 for a total time of 
14\,h. Positive excesses have been 
observed at low significances of
1.6\,$\sigma$~(2000) and 0.9\,$\sigma$ (2001) leading to a 3\,$\sigma$
upper limit of
$N_\gamma(E > 250\,\mbox{GeV}) < 2.2 \times 10^{-11}$
phot.\,cm$^{-2}$s$^{-1}$ 
(\cite{m87_whipple_icrc_2001}).

The HEGRA collaboration has extensively observed \mmm
in 1998 and 1999 with the stereoscopic system 
of 5~imaging atmospheric Cherenkov telescopes (IACT system, \cite{hegra_iact_system_1997}).
About half of the total observation time (44.1\,h out of 83.4\,h) has been used
in an earlier analysis (\cite{hegra_m87_icrc_2001}). 
No evidence for TeV emission was found in this dataset 
and a 3\,$\sigma$ upper limit on the TeV $\gamma$-ray flux from \mmm was
determined to be
$N_\gamma(E > 720\,\mbox{GeV}) < 1.45 \times 10^{-12}$\,phot.\,cm$^{-2}$s$^{-1}$.

In this Letter the results of the whole data set of the extensive HEGRA \mmm observations
during the years 1998 and 1999 are reported, now also applying
a more sensitive analysis method.
Astrophysical conclusions concerning the nature of the observed
excess are briefly discussed.


\section{Observations and results of analysis}

\mmm was observed in the years 1998 and 1999
with the HEGRA IACT system for a total of 83.4\,h.
There were no further observations of \mmm with the HEGRA telescopes in the
subsequent years.
The major part of the \mmm data was taken with a 4-telescope setup.
%
Table~\ref{hegra_observations} specifies the observation times and mean 
zenith angles of the individual HEGRA observation periods. The mean zenith angle
of~21.6$^\circ$ can be converted into a mean energy threshold (defined as the
peak detection rate for $\gamma$-showers) 
of 730\,GeV for a Crab-like spectrum (\cite{Konopelko_1999}).
\begin{table}[t]
    \begin{center}
      \begin{tabular}{lc|crcc} \hline \hline
        Date                     &     Year   &   \multicolumn{3}{c}{Obs. Time}   &     $\langle \vartheta \rangle$ \\
                                 &            &   \multicolumn{3}{c}{[h]}         &     [$^\circ$] \\
        \hline
        December 28              &   1998     &  & 0.7&        &   23.6   \\
        Jan.~17 -- Jan.~26       &   1999     &  &10.7&        &   18.4   \\
        February 12              &   1999     &  & 0.7&        &   17.0   \\
        March 16 -- March 24     &   1999     &  &21.7&        &   21.4   \\
        April 5 -- April 21      &   1999     &  &29.4&        &   23.2   \\
        May 8 -- May 18          &   1999     &  &19.9&        &   20.9   \\
        June 3                   &   1999     &  & 0.3&        &   40.5   \\ \hline
        Total                    &            &  &83.4&        &   21.6   \\ \hline
      \end{tabular}
    \end{center}
    \caption{\label{hegra_observations}
      Dates of individual HEGRA observation periods of \mmm. Listed are
      observation times and mean zenith angles $\langle \vartheta
      \rangle$. 
      Typically, each night comprises approx.~1 -- 2\,h of observation time.}
\end{table}
\begin{table}[ht]
    \begin{center}
      \begin{tabular}{lll} \hline \hline
            &  \multicolumn{2}{l}{\mmm event selection}  \\ 
        \hline
        stereo algorithm                           &     \#3                 \\
        number of images per event \quad           &     $\ge$ 2             \\
        shape cut on~~{\em mscw}                   &     $<$ 1.1             \\
        angular distance cut~~$\Theta^2 $          &     $<$ 0.016 deg$^2$   \\
        \hline
           &   \multicolumn{2}{l}{background models:}                \\
        %
        %
            &  \underline{\em ring segment}        &  \underline{\em template} \\
        $N_{\mbox{\tiny ON}}$                      &      716               &  716   \\
        $N_{\mbox{\tiny OFF}}$                     &     6950               &  1850  \\
        $\alpha = \Omega_{\mbox{\tiny ON}} / \Omega_{\mbox{\tiny OFF}}$ & 0.08757     &  0.325 \\
        $N_{\mbox{\tiny $\gamma$-candidates}}$     &      107.4 $\pm$ 26.8  &  114.8 $\pm$ 30.2 \\
        significance ($\sigma$)                    &        4.1             &  3.9   \\
        \hline
      \end{tabular}
    \end{center}
    \caption{\label{hegra_numbers}
      Cuts, event numbers, and significances for the HEGRA observations
      of \mmm resulting from the signal search using the ring segment and
      template background model, respectively (see text).}
\end{table}

Only data of good quality were considered for the analysis, the most 
critical condition being the IACT system's cosmic ray background
trigger rate not deviating more than 30\,\% from the rate expected
for the current zenith angle. A total of 
about 5\,\% of the data was rejected due to this selection.

All observations of \mmm were carried out in the so-called 
{\em wobble} mode targeting the object's position (``ON'') as given in 
Section \ref{chapter1}
shifted by $\pm$0.5$^\circ$ in declination with respect to the center of 
the field of view. This observation mode allows for simultaneous estimation of the 
background (``OFF'') rate induced by charged cosmic rays
(\cite{Aharonian_1997}). 
The analysis uses an extended OFF-region reducing the statistical error
on the number of background events. A ring segment is chosen with 260$^\circ$
opening angle at the same radial distance to the center of the 
field of view as for the ON-source position (see also \cite{hegra_mrk-421_2002}). 
The width of the ring is set to the diameter of the ON-source area in order to 
provide the same angular acceptance for ON- resp.~OFF-source events. This ring 
segment background model is similar to the usage of a set of control regions (\cite{hegra_cas-a}),
but provides a smaller ratio of ON- to OFF-source solid angle areas
($\alpha = \Omega_{\mbox{\tiny ON}}/\Omega_{\mbox{\tiny OFF}}$) and thus reduces the 
statistical error on the number of estimated background events.
%
For a consistency check (and for a search for
$\gamma$-ray sources in the field of view, see below) the so-called template
background model has also been used (\cite{template_model}, see also
\cite{hegra_cyg-ob2}).

For the image analysis, the mirror reflectivities and photocathode
efficiencies -- which degrade slowly with time -- along with
the factors converting from digitized photomultiplier
signals to photoelectrons have been determined on a monthly basis. 
The shower reconstruction and the event selection cuts have already been 
described in previous publications (e.\,g.~\cite{Aharonian_1999}). 
The stereo air shower direction
reconstruction algorithm~\#3 (\cite{Hofmann_1999}) and a ``tight shape cut'' 
(parameter {\em mscw}~$< 1.1$) (\cite{Konopelko_1999}) for an effective
$\gamma$-hadron separation have been applied leading to a sensitivity
gain as compared to the earlier analysis of the HEGRA \mmm observations.
The optimum angular cut was derived using $\gamma$-ray
events from the Crab nebula on the basis of a nearly contemporaneous data set
at similar zenith angles. 
Table~\ref{hegra_numbers} summarizes the event selection cuts, the resulting event 
numbers and significances for the data set as derived using the
ring segment and template background models.

\begin{figure}[t]
  \centering
  \includegraphics[width=8.5cm]{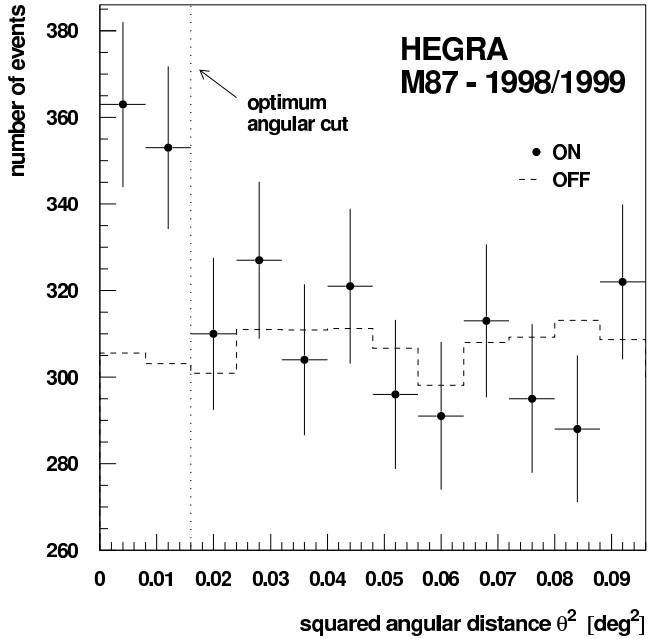}
  \caption{
    Number of events vs.~squared angular distance $\Theta^2$ to the position of
    \mmm as observed in the years 1998 and 1999 with the HEGRA IACT system. 
    The dots show the ON-source events, while the dashed histogram
    gives the background estimate determined from a ring segment
    as explained in the text. The statistical error for
    the background estimate is much smaller than the error of the 
    ON-source distribution. Indicated by the vertical dotted line is
    the optimum angular cut as determined from nearly contemporaneous Crab
    observations at similar zenith angles. The significance of the \mmm excess 
    amounts to 4.1\,$\sigma$.
    }
  \label{M87_onoff}
\end{figure}

Figure~\ref{M87_onoff} shows the event distribution both 
for the ON-source and the OFF-source regions as a function of the squared
angular distance of the reconstructed shower direction to the source position
after applying all event selection cuts. 
The statistical significance of the observed excess from the direction 
of \mmm (at the reference coordinates given in Section \ref{chapter1}) 
is~4.1\,$\sigma$, calculated using formula (17) from \cite{Li_Ma}.
On the basis of the limited event statistics the excess is compatible 
with a point-like source for the HEGRA IACT system at the position of \mmm, 
although extended emission cannot be excluded. 
After background subtraction the {\em mscw} values show a Gaussian 
distribution around the value of 1.0 as expected for a $\gamma$-ray 
population (see \cite{Konopelko_1999}). This test supports the hypothesis that 
the measured excess is a result of \mmm being a true TeV $\gamma$-ray source.
%
%
%
%
%
Applying different statistical tests in order to search for burstlike
behaviour of \mmm no evidence for flux variations in the TeV energy range has been found.

\begin{figure}
  \centering
  \includegraphics[bbllx=70, bblly=147, bburx=439, bbury=498, clip=, width=8.5cm]{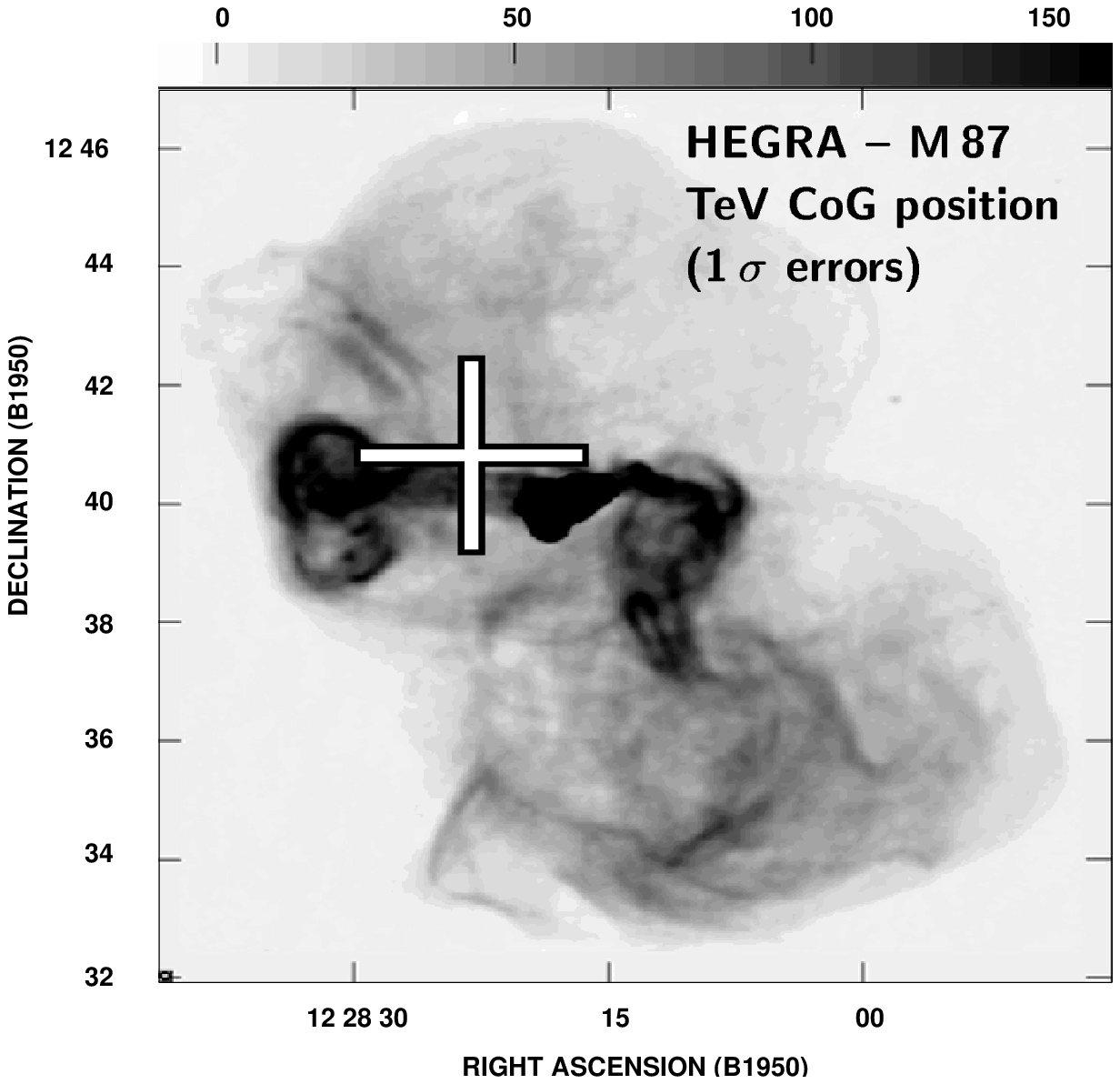}
  \caption{
    Radio image of \mmm at 90\,cm showing the structure of the 
    \mmm halo.
    The center of gravity position of the TeV $\gamma$-ray excess
    from the HEGRA \mmm observations is marked by the cross 
    indicating the statistical 1\,$\sigma$ errors 
    (radio image adapted from \cite{m87_owen_2000b}).}
  \label{M87_radio_tev}
\end{figure}

The event distribution in the field of view was used 
to determine the center of gravity position (CoG) of the 
TeV $\gamma$-ray excess at 
$\alpha_{\mbox{\tiny J2000.0}} = 12^{\mbox{\tiny hr}}30^{\mbox{\tiny m}}54.4^{\mbox{\tiny s}}
  \pm 6.9^{\mbox{\tiny s}}_{\mbox{\tiny stat}} \pm
  1.7^{\mbox{\tiny s}}_{\mbox{\tiny syst}}$,
$\delta_{\mbox{\tiny J2000.0}} = 12^\circ24^{\prime}17^{\prime\prime}
  \pm 1.7^{\prime}_{\mbox{\tiny stat}} \pm 0.4^{\prime}_{\mbox{\tiny syst}}$
as shown in Figure \ref{M87_radio_tev}.
The accuracy of the CoG determination is limited by a systematic
pointing error of about 25\,$^{\prime\prime}$ (\cite{puehlhofer_1997}).
Within the large errors, the CoG is consistent with the \mmm position,
although a small shift of the source position cannot be ruled out.
Therefore, it is not possible to localize a candidate TeV
$\gamma$-ray production site to particlular inner radio structures of \mmm.

\begin{figure}[ht]
  \centerline{\includegraphics[width=8.8cm]{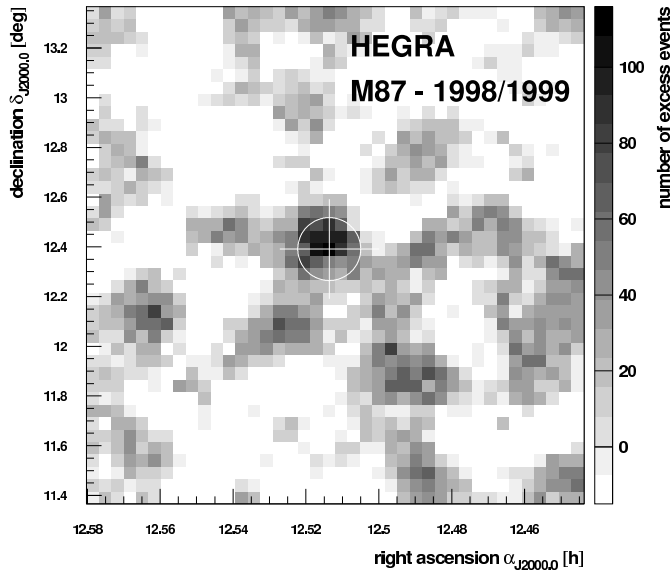}}
  \caption{
    Skymap ($2^\circ \times 2^\circ$ view at a $0.02^\circ \times 0.02^\circ$
    binning) of the number of excess events in the HEGRA field of view around the position of \mmm generated
    using the template background model. At each bin, the excess is determined
    from events within the optimum radius of $0.127^\circ$ 
    (based on Crab observations). Negative excesses below -15 events are shown
    in white color for clarity reasons.
    The solid angle resulting from the optimum radius is indicated by the white circle
    with \mmm being located at the intersection of the white lines.
    }
  \label{M87_skymap}
\end{figure}

In order to search for TeV $\gamma$-ray sources in the relatively large 
field of view of the HEGRA IACT system (and for a consistency check with the
ring segment background model), the template background 
model was used for 
a further analysis of the \mmm data. A $2^\circ \times 2^\circ$ skymap
of excess events determined for this sky region
is shown in Figure \ref{M87_skymap}. The excess from the direction of \mmm is clearly
visible in the representation using overlapping circular bins, showing
that the only significant excess in the field of view is related
to \mmm. The significance using the template model is $3.9\,\sigma$ as given in
Table~\ref{hegra_numbers}. The two background models applied in this 
analysis use widely different approaches, supporting
the assumption that the \mmm excess does not stem from a background fluctuation.

The observed excess can be converted into an integral flux 
of ($3.3 \pm 0.8)$\,\% of the Crab nebula flux
(only the statistical error is given because of the low statistics).
A conversion into absolute flux units 
using the well measured photon flux and spectrum of the Crab nebula around
1\,TeV (e.\,g.~\cite{hegra_crab})
results in a $\gamma$-ray flux of
\begin{equation}
  N_\gamma(E > 730\,\mbox{GeV}) = (0.96 \pm 0.23) \times 10^{-12}\,\mbox{phot.\,cm}^{-2}\,\mbox{s}^{-1}\mbox{.}
\end{equation}
A spectral analysis of the data of the \mmm data has been performed using
the analysis technique described in \cite{hegra_h1426_paper2}.
The data can be well described with a power law 
$\mbox{d}N/\mbox{d}E \sim E^{-\alpha}$ with
$\alpha = 2.9 \pm 0.8_{\mbox{\tiny stat}} \pm 0.08_{\mbox{\tiny syst}}$. 
The large statistical error results from the low event statistics.


\section{Summary and Conclusions}

The radio galaxy \mmm has been observed with the HEGRA IACT system for a total of~83.4\,h. 
For the first time a significant excess of 
4.1\,$\sigma$ has been detected at energies above a mean energy threshold of 730\,GeV
from a member of this class of objects using the imaging atmospheric Cherenkov
technique. 
Note that nearly 30 years ago the radio galaxy Centaurus~A was reported to be a TeV source 
observed with a non-imaging instrument (\cite{grindlay}).
%

Due to the limited number of excess events detected by the HEGRA telescopes, an analysis of the spectral shape 
results in a very large statistical error, thus making it difficult to draw 
a conclusion about the origin of
the TeV $\gamma$-radiation. Assuming a power-law shape with a photon spectral
index of 2.9 the integral photon flux at the level of ($3.3 \pm 0.8)$\,\% of
the Crab nebula flux converts into an energy flux of
\begin{equation}
  F_\gamma(E > 730\,\mbox{GeV}) = (2.4 \pm 0.6) \times 10^{-12}\,\mbox{erg\,cm}^{-2}\,\mbox{s}^{-1}\mbox{.}
\end{equation}
Given the distance to \mmm of about 16\,Mpc, this
corresponds to a $\gamma$-ray luminosity above 730\,GeV 
of about $10^{41}\,\mbox{erg\,s}^{-1}$ under the assumption of isotropic emission.

Several different possibilites for the origin of GeV/TeV $\gamma$-radiation 
are conceivable.
\mmm with its pc scale jet has recently been modeled
within the Synchrotron Self Compton scenario
as a BL Lac object seen at a large angle to its jet axis (\cite{Bai_Lee}).
It has also been modeled using the so-called Synchrotron Proton Blazar
model (\cite{m87_protheroe}). In both models, the observed flux can
be accommodated.

The large scale (kpc) jets with several knots detected at radio to X-ray 
frequencies and believed to have synchrotron origin due to electrons with 
energies up to 100\,TeV is also a possible $\gamma$-ray production site in \mmm.
Consequently, inverse Compton $\gamma$-rays in the 1 -- 10\,TeV
energy range can be expected within reasonable model parameters.

Moreover, $\gamma$-rays could be produced in the interstellar medium
of \mmm, i.\,e.~at larger distance scales from the center of this active galaxy. 
Both inverse Compton and hadronic interactions could generate a TeV
$\gamma$-ray luminosity in the observed range of 
approx.~$10^{41}\,\mbox{erg\,s}^{-1}$. If this interpretation
is valid, one would expect a slightly extended source (2 --
3$^{\prime}$).

It should be noted that M\,87 is also considered as a possible source of TeV $\gamma$-rays from
the hypothetical neutralino annihilation process (\cite{baltz_neutralino}).

A weak signal at the centi-Crab level is at the sensitivity threshold
for the HEGRA IACT system for observation times of the order of 100\,h.
Therefore, a deep investigation and possible confirmation with a spectral
analysis of the \mmm
excess should be subject of the next generation Cherenkov telescope projects like
\hess, MAGIC and VERITAS, 
 which provide increased sensitivity together with a lower energy threshold. 
Due to the proximity to \mmm (16\,Mpc) and to the increased accuracy of
these observations (a fraction of an arc minute), these measurements may allow 
the location of the $\gamma$-ray production site in \mmm
to be more accurately determined
thus greatly advancing our understanding of its TeV $\gamma$-radiation.
%


\begin{acknowledgements}

  The support of the German Federal Ministry for Research and Technology BMBF and
  of the Spanish Research Council CICYT is gratefully acknowledged. 
  G.~Rowell acknowledges receipt of a von Humboldt fellowship.
  We thank the Instituto de Astrof\'{\i}sica de Canarias (IAC)
  for the use of the HEGRA site at the Observatorio del Roque de los 
  Muchachos (ORM) and for supplying excellent working conditions on La
  Palma.

\end{acknowledgements}


\end{document}